\documentclass{ws-p8-50x6-00}

\begin{document}

\title{Thermalization and elliptic flow at RHIC}

\author{Nicolas Borghini}
\address{Service de Physique Th\'eorique, CP225, 
Universit\'e Libre de Bruxelles, 1050 Brussels, Belgium}

\author{Phuong Mai Dinh and \underline{Jean-Yves Ollitrault}}

\address{Service de Physique Th\'eorique, CEA-Saclay, 
91191 Gif-sur-Yvette cedex, France}
\maketitle

\abstracts{
We discuss signatures of thermalisation in heavy ion collisions 
based on elliptic flow. 
We then  propose a new method to analyse elliptic flow, 
based on multiparticle azimuthal correlations. This 
method allows one to test quantitatively the collective behaviour 
of the interacting system.}

\section{Introduction}

A highly debated issue in ultrarelativistic heavy ion 
collisions is whether or not the colliding system reaches 
local thermal equilibrium at some stage of its evolution. 
Equilibrium with 
respect to inelastic collisions strongly constrains 
the ratios of particle abundances~\cite{Becattini:1998ii}
(``chemical'' equilibrium), as well as phase-space 
densities, which are obtained by combining informations from 
momentum spectra and two-particle HBT 
correlations;\cite{Bertsch:1994qc} 
on the other hand, equilibrium with respect to elastic collisions 
constrains momentum distributions, and implies in 
particular that they are isotropic in the local 
rest frame. This is the ``kinetic'' equilibrium, on 
which we concentrate here. 

Kinetic equilibrium itself has (at least) two facets.
One is the equilibration between longitudinal and transverse
degrees of freedom, i.e., the implication that in the local 
rest frame, longitudinal and transverse momenta are of 
the same order of magnitude. 
This aspect of thermalisation can be discussed from first 
principles at the partonic level,\cite{Shuryak:1992wc}
and there is now a vast literature on this subject.\cite{Baier:2001sb} 
But experimental signatures deal in fact rather with 
equilibration among the two transverse degrees of freedom.
Due to the high Lorentz contraction at ultrarelativistic 
energies, the typical transverse scale is much larger 
than the longitudinal scale, so that this 
``transverse equilibrium'' is probably easier to achieve than 
``longitudinal-transverse equilibrium''. 

In Sec.~\ref{sec:hydro}, we discuss some experimental signatures 
of transverse equilibrium, especially elliptic flow, 
and the relevance of hydrodynamical models in this context. 
In Sec.~\ref{sec:analysis}, we present a new method recently developed
to obtain more reliable measurements of elliptic flow,
which are required in order to draw definite conclusions on the 
issue of thermalisation. 

\section{Elliptic flow: a signature of transverse thermalisation}
\label{sec:hydro}

\subsection{When are hydrodynamical models useful?}

If the colliding system thermalises, its evolution follows the 
laws of hydrodynamics; it is therefore natural to use hydrodynamical 
models in order to define signatures of thermalisation. 
There are two ingredients in these models:
i) an equation of state; 
ii) initial conditions, i.e., energy density, baryon density 
and fluid velocity on a space-like hypersurface (typically, at 
some initial time).

Which parameters are under control?
The initial time $t_0$, at which the system thermalises, is hopefully 
short but to a large extent unknown, so that hydrodynamics has 
little predictive power concerning observables which depend strongly 
on $t_0$, such as thermal photon 
production.\cite{Srivastava:2001pv} 
Similarly, there is no serious motivation for studying the 
very first stages of the collision within the framework 
of hydrodynamics, either within a one-fluid 
model~\cite{Hung:1995eq,Rischke:1995pe} or a multifluid 
model,\cite{Brachmann:2000mp} and the relevance of 
signatures based on such parametrisations is questionable. 
Finally, there is no well-defined prescription concerning the 
initial longitudinal fluid velocity and density, for which 
a variety of parametrisations exist, either 
``Landau''~\cite{Bolz:1992nt} or ``Bjorken''~\cite{Bjorken:1983qr} 
type. 

The situation is much clearer concerning transverse degrees 
of freedom: the transverse collective velocity must be 
initially zero since each nucleon-nucleon collision populates 
the transverse momentum space randomly; 
the initial density profile in the transverse plane
is strongly constrained by observed multiplicity 
distributions.\cite{Ollitrault:1991xx} 
Therefore, reliable signatures of thermalisation should 
rather be sought for in observables associated with transverse
degrees of freedom: $p_T$ spectra,\cite{Schnedermann:1993ws}
transverse radii~\cite{Appelshauser:1998rr}
and azimuthal anisotropies, in particular elliptic flow, 
on which we concentrate here. 

\subsection{Why elliptic flow?}

Elliptic flow is defined as a correlation between the 
azimuth $\phi$ of an outgoing particle and the azimuth 
$\Phi_R$ of impact parameter:\cite{Voloshin:1996mz} 
\begin{equation}
\label{eq:defv2}
v_2=\left<e^{2i(\phi-\Phi_R)}\right>,
\end{equation}
where brackets denote a statistical average.
At ultrarelativistic energies, $v_2$ is positive 
for noncentral collisions.\cite{Barrette:1997pt,Appelshauser:1998dg}
In a hydrodynamical picture, it results from anisotropic 
pressure gradients in the transverse plane, due to the 
almond-shaped region of the overlap region between the 
two nuclei.\cite{Ollitrault:1992bk}  
Microscopically, $v_2$ is created by rescattering among the 
produced particles, which makes it a sensitive probe 
of final state interactions: if there are none, 
it vanishes, while other observables such as $p_T$ spectra 
may still look ``thermal''.\cite{Schaffner-Bielich:2001qj}

Furthermore, predictions of hydrodynamical models for 
$v_2$ are very stable:
since $v_2$ is created by transverse pressure gradients, 
it strongly depends on the initial density transverse profile,
which is well controlled as discussed above; 
it also depends significantly on the equation of state, on 
which it may thus provide valuable information;
on the other hand, $v_2$ depends only weakly on arbitrary 
parameters, such as initial time and longitudinal 
velocity.\cite{Ollitrault:1992bk} 
Quite remarkably, simple hydrodynamical parametrisations 
are able to reproduce simultaneously the measured $p_T$ spectra, 
HBT radii and elliptic flow at RHIC.\cite{Raimond} 

\subsection{Predictions for $v_2$}

The momentum anisotropy $v_2$ calculated in hydro 
models is roughly equal to the anisotropy of the 
almond-shaped overlap region.\cite{Ollitrault:1992bk}  
This purely geometrical effect dominates the centrality 
dependence of $v_2$, which decreases linearly with 
the number of participants. 
Deviations from this behaviour can be used to signal 
a phase transition~\cite{Sorge:1999mk} or 
a departure from thermal equilibrium.\cite{Voloshin:2000gs}
The latter is expected to occur for the most peripheral collisions,
where $v_2$ should be smaller than the hydro prediction. 
This is indeed observed in several specific transport models 
like UrQMD~\cite{Bleicher:2000sx} (which however predicts 
a much too small value of $v_2$), QGSM,\cite{Zabrodin:2001rz}
and AMPT.\cite{Lin:2001zk}
With a more systematic study,\cite{Molnar:2001ux} one could 
relate the observed centrality dependence of $v_2$ to the 
degree of thermalisation of the system. 
Experimental results vary significantly depending on the method 
used to analyse elliptic flow.\cite{Ackermann:2001tr,Tang:2001yq}
We come back to this issue in Sec.~\ref{sec:analysis}.

Hydrodynamical calculations~\cite{Kolb:2000sd} were
also able to predict the $p_T$ dependence of $v_2$ for 
identified hadrons, in remarkable agreement with 
experimental results:\cite{Adler:2001nb}
$v_2$ is almost linear in $p_T$ for pions and significantly 
smaller for protons. However, these non-trivial features 
are also reproduced by transport 
models.\cite{Bleicher:2000sx,Zabrodin:2001rz,Lin:2001zk}
In addition, the latter predict a saturation~\cite{Molnar:2001ux} 
at high $p_T$ which is not seen in hydro, suggesting 
that many elastic collisions are necessary to build the 
flow at high $p_T$.
This saturation, which is seen in the data, 
could also be related to hard physics.\cite{Wang:2001fq}

\section{Analysing elliptic flow with multiparticle correlations}
\label{sec:analysis}

\subsection{Flow from azimuthal correlations} 

Since the reaction plane $\Phi_R$ is unknown, 
$v_2$ cannot be derived directly from (\ref{eq:defv2}). 
It must be inferred from azimuthal correlations. 
The standard flow analysis~\cite{Danielewicz:1985hn} relies 
on the key assumption that all azimuthal 
correlations are due to flow, i.e., that 
angles relative to the reaction plane $\phi-\Phi_R$ 
are statistically independent. 
This allows one to write the two-particle 
correlation as~\cite{Wang:1991qh}
\begin{equation}
\label{eq:twoflow}
\left< e^{2i(\phi_1-\phi_2)}\right>= 
\left< e^{2i(\phi_1-\phi_R)} e^{2i(\phi_R-\phi_2)}\right>= 
 (v_2)^2.
\end{equation}
where brackets denote an average over pairs of particles
belonging to the same event. 
One could also use higher order correlations, such 
as the four-particle correlation, which would give
\begin{equation}
\label{eq:fourflow}
\left< e^{2i(\phi_1+\phi_2-\phi_3-\phi_4)}\right>
= (v_2)^4. 
\end{equation}
However, such equations are not quite correct, since they neglect
various other sources of azimuthal correlations (``nonflow correlations''), 
which is no longer justified at ultrarelativistic 
energies.\cite{Dinh:2000mn}

\subsection{Simple illustration of nonflow correlations}

In order to illustrate nonflow correlations, we consider 
the following example: assume that in each event, $M/2$ pairs 
of particles are emitted, where both particles in a pair 
have collinear momenta, but pairs are emitted with random 
orientations. 
Since pairs are emitted randomly, there is no flow ($v_2=0$), 
but there are azimuthal correlations. 
In each event, there is a total of $M(M-1)/2$  
particle pairs, among which $M/2$ are correlated, 
hence the two-particle correlation 
\begin{equation}
\label{eq:twonf}
\left< e^{2i(\phi_1-\phi_2)}\right>= {1\over M-1}.
\end{equation}
A similar reasoning yields the four-particle correlation:
\begin{equation}
\label{eq:fournf}
\left< e^{2i(\phi_1+\phi_2-\phi_3-\phi_4)}\right>
= {2M(M-2)\over M(M-1)(M-2)(M-3)}={2\over (M-1)(M-3)}.
\end{equation}
Applying Eqs.\ (\ref{eq:twoflow}) and (\ref{eq:twonf}), 
or (\ref{eq:fourflow}) and (\ref{eq:fournf}), 
one would obtain $v_2\sim 1/\sqrt{M}$, although there 
is no flow: this is the typical order at which 
nonflow correlations spoil the standard flow analysis. 

\subsection{Subtraction of nonflow correlations}

The contribution of nonflow correlations can be 
greatly reduced by combining 
the informations from two- and four-particle correlations. 
In Eq.\ (\ref{eq:fournf}), a 4-uplet of particles gives 
a nonvanishing contribution if it consists of two 
correlated pairs, either (1,3) and (2,4) or (1,4) 
and (2,3). Subtracting the corresponding contributions, 
one obtains
\begin{equation}
\label{eq:fourcumulant}
\left< e^{2i(\phi_1+\phi_2-\phi_3-\phi_4)}\right>
-2\left< e^{2i(\phi_1-\phi_2)}\right>^2
= {4\over (M-1)^2(M-3)}.
\end{equation} 
The l.-h.\ s.\ of this equation defines the {\it cumulant\/} 
of four-particle correlations, $c_2\{4\}$. 
The r.-h.\ s.\ is the contribution of nonflow correlations,
of order $1/M^3$, i.e., much smaller than the 
corresponding contribution to the four-particle correlation 
(\ref{eq:fournf}), of order $1/M^2$. 
On the other hand, the contribution of flow remains of the 
same magnitude: from Eqs.\ (\ref{eq:twoflow}), (\ref{eq:fourflow})
and (\ref{eq:fourcumulant}), one obtains $c_2\{4\}=-(v_2)^4$. 
The subtraction therefore reduces the relative contribution 
of nonflow effects, and yields a more accurate estimate of the 
flow. 

This method was recently applied to STAR data.\cite{Tang:2001yq}
The resulting value of $v_2$ are smaller than those obtained with the 
standard analysis,\cite{Ackermann:2001tr}
in particular for the most peripheral collisions:
this is precisely where nonflow effects are expected to give 
the largest contribution since the multiplicity $M$ is smaller.  
The centrality dependence obtained with this method suggests 
that departures from thermalisation at RHIC may be larger than 
was previously thought. 

This cumulant expansion can be worked out 
to arbitrary orders, and allows one to extract the genuine 
4-, 6-particle correlations and beyond.\cite{Borghini:2001sa} 
The practical implementation of the method is described in 
detail elsewhere.\cite{Borghini:2001zr} 
Flow, which is essentially a collective phenomenon, 
contributes to all orders, while the relative contribution 
of nonflow correlations decreases as the order increases. 
Higher order cumulants therefore 
provide a unique possibility to check quantitatively 
that azimuthal correlations are indeed of collective origin.

\section*{Acknowledgments}
J.-Y. O. thanks Raimond Snellings and Raju Venugopalan 
for discussions.

\end{document}